\documentclass[aps,prl,twocolumn,10pt,superscriptaddress,floatfix,noeprint,nofootinbib]{revtex4-2}
\usepackage[T1]{fontenc}
\usepackage[latin9]{inputenc}
\usepackage{mathrsfs}
\usepackage{leftindex}
\usepackage[caption=false]{subfig}
\usepackage{amsmath}
\usepackage{amssymb}
\usepackage{graphicx}
\usepackage{hyperref}
\usepackage[export]{adjustbox}
\hypersetup{colorlinks = true,
	linkcolor =blue,
	citecolor=blue, 
	urlcolor=blue 
}
\usepackage[normalem]{ulem}
\usepackage{xcolor}

\newcommand{\soutcolor}[2][red]{\bgroup\markoverwith{\textcolor{#1}{\rule[0.5ex]{2pt}{0.4pt}}}\ULon{#2}\egroup}

\usepackage{overpic}
\usepackage{xcolor}

\newcommand{\trajavg}[1]{\langle\langle #1 \rangle\rangle}
\newcommand{\ket}[1]{\ensuremath{\left\vert #1 \right\rangle}}
\newcommand{\bra}[1]{\ensuremath{\left\langle #1 \right\vert}}
\newcommand{\mr}[1]{\ensuremath{\mathrm{#1}}}

\newcommand{\hrho}{\ensuremath{\hat{\rho}}}

\newcommand{\Diss}[2]{\ensuremath{\mathcal{D}(#1,#2)}}

\newcommand{\Hop}{\hat{H}}

\newcommand{\rhoop}{\hat{\rho}}

\newcommand{\sigop}{\hat{\sigma}}
\newcommand{\hchi}{\hat{\chi}}
\usepackage{bm}

\usepackage[english]{babel}

\begin{document}
\title{Measurement Induced Subradiance}
\author{Ipsita Bar}
\email{ipsitab@iitgn.ac.in }
\affiliation{Indian Institute of Technology Gandhinagar, Palaj, Gujarat 382355, India}

\author{Aditi Thakar }
\email{aditi.thakar@alumni.iitgn.ac.in}
\affiliation{Indian Institute of Technology Gandhinagar, Palaj, Gujarat 382355, India}
\author {B. Prasanna Venkatesh}
\email{prasanna.b@iitgn.ac.in}
\affiliation{Indian Institute of Technology Gandhinagar, Palaj, Gujarat 382355, India}
\begin{abstract}
Preparing subradiant steady states of collectively emitting quantum two-level emitters (TLEs) is hindered by their dark, weakly interacting nature. Existing approaches rely on patterned driving, local control, or structured environments. We propose a platform-independent protocol based on projective measurements on a single TLE. For permutation-symmetric ensembles, a single measurement yields appreciable occupation of single-excitation subradiant steady states. For generic arrays, repeated measurements on one emitter drive the unmeasured TLEs into a nearly pure state with large overlap with the subradiant Dicke subspace.

\end{abstract}
\maketitle
\noindent\textit{Introduction.}---Subradiance, the suppression of collective spontaneous emission in ensembles of two-level emitters (TLEs), is a paradigmatic manifestation of cooperative light-matter interactions \cite{ClaudiuReviewPRXQuantum_2022,ChangCollectiveNanoLatticeReview2018}. The resulting long-lived collective states offer a promising resource for robust entanglement generation, quantum sensing, quantum memories, and decoherence-free subspaces \cite{Santos2022_GenEntanglement,Santos2023_GenEntanglement,OstermannPRL2013,OstermannPRA2014,zafrabono2025subradiantcollectivestatesprecision,Facchinetti2016,AsenjoGarcia2017,Paulisch_2016}. While transient occupation of subradiant states has been achieved across multiple platforms such as ion traps, quantum dots, ultracold atoms, and superconducting qubits \cite{DeVoe1996SuperSubradiantTwoIons,Tiranov2023AtomicArraySubradiance,Pellegrino2014SuppressionScattering,Guerin2016ColdAtomSubradiance,Solano2017SuperradianceInfiniteRangeNC,Rui2020SubradiantOpticalMirror,Ferioli2021,Mirhosseini2018SCMetamaterialsWQED,Kannan2021WaveguideBandgapSCArray,Zanner2022DarkStateWQED,Sharafiev2025CollectiveEffectsSC}, the controlled preparation of subradiant steady-states remains challenging. Existing approaches typically rely on engineered photonic environments \cite{vanLoo2013PhotonMediatedArtificialAtoms,Wang2020SCQubitSubradiance,Yan2023SuperradiantSubradiantCavity,Gabor2025SubradiantAtomArrayEPJQT,BaumgartnerDonner2025,kim_cavity-mediated_2025}, spatially patterned driving \cite{plankensteiner_selective_2015,manzoni_optimization_2018,guimond_subradiant_2019,RubiesBigorda2022,Fayard2023}, ancillary systems \cite{Albrecht2019}, or individual phase control of emitters \cite{Filipp2011,jenkins_controlled_2012,jen_cooperative_2016}, which become increasingly demanding for large arrays. A simple and platform-independent mechanism for preparing subradiant steady-states is therefore still lacking.

\begin{figure}
\centering
\begin{overpic}[width=\linewidth]{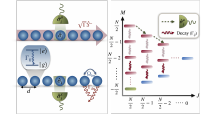}\put(1,52){\textbf{(a)}} \put(51,52){\textbf{(b)}}
\end{overpic}
\caption{(a) Schematic of $N$ two-level emitters undergoing collective spontaneous emission and subject to single emitter (local) measurements. Top panel depicts a permutation symmetric ensemble (PSE) subjected to a projective measurement of $\sigop_i^{\mu=(x,z)}$ (shown in green) and bottom panel shows a generic array where repeated measurements of $\sigop_i^{x}$ are applied. (b) Population dynamics in the Dicke basis $\vert J,M\rangle$: In a PSE, collective decay (red curly arrows) redistributes population across Dicke ladders with the same total spin $J$ but local measurements (dashed green arrows) provide an irreversible way to  
redistribute population across different $J$ ladders, steering the system away from superradiant to subradiant states (blue).}
\label{fig:Schematic}
\end{figure}

Quantum measurement back-action provides a markedly different control method, allowing probabilistic but scalable state engineering without tailored hamiltonian engineering and enables a range of quantum information processing tasks \cite{Briegel,MeasQuantComputeExpt2013,Kuzmich1998QNDSpinSqueezing,Thomsen2002,Appel2009,hosten_measurement_2016,Albarelli_2017,Genoni2020,guerlin_progressive_2007,sayrin_real-time_2011,ShimizuMeasCAT2018,nosrati_robust_2020,GefenSteering2020A,GefenSteering2020B,GefenSteering2023A,GefenSteering2023B,GefenSteering2024A,GefenSteering2024B,Piroli2024,OConnor_2025,LoFrancoGenIdentity2024,Puente2024quantumstate}. In collectively dissipative systems, measurement based protocols have thus far only involved \emph{collective} measurements to prepare superradiant Dicke states \cite{stockton_deterministic_2004,YuEffBrightDicke2026} or subradiant states \cite{ganesh_generating_2017}. Here, we show that local quantum measurements alone suffice to engineer subradiant states in emitter arrays undergoing collective spontaneous emission. Specifically, we demonstrate that measurements performed on a single emitter [see Fig.~\ref{fig:Schematic} (a)] at appropriate times can steer the system into subradiant Dicke states in permutation symmetric ensembles (PSEs). Furthermore, in generic arrays of $N$ TLEs lacking permutation symmetry, repeated measurements on one emitter lead to the preparation of subradiant steady states of the remaining unmeasured $N-1$ emitters. Our scheme requires neither spatially patterned drives nor structured reservoirs, and is directly applicable across diverse experimental platforms, thereby establishing local measurement back-action as a minimal resource for generating subradiant many-body states.

To understand the central idea of our proposal consider the collective spontaneous emission of two emitters initially in the excited state $\ket{ee}$ described using the Monte-Carlo wavefunction (MCWF) quantum trajectory evolution formalism \cite{Dalibard1992MCWF,Carmichael1993OpenSystems,Molmer1993MCWF,Molmer1996ReviewMCWF}. Following the emission of the first photon, the system is projected to the symmetric or bright Dicke state $(\ket{eg} + \ket{ge})/\sqrt{2}$. A local measurement of the excitation of one of the emitters, while in this transient state before it reaches the ground state $\ket{gg}$, projects the system to $\ket{eg}$ or $\ket{ge}$. These states are an equal superposition of the bright and the dark Dicke state ($(\ket{eg} -\ket{ge})/\sqrt{2}$) which naturally guarantees that a finite fraction of quantum trajectories subject to the measurement described above will terminate in a subradiant Dicke state. This idea can be further generalized to multiple emitters as illustrated in Fig.~\ref{fig:Schematic} (b) where local quantum measurements provide a way to couple the ladder of bright Dicke states of $N$ TLEs to the subradiant Dicke states.

\begin{figure}
\centering
\begin{overpic}[width=\linewidth]{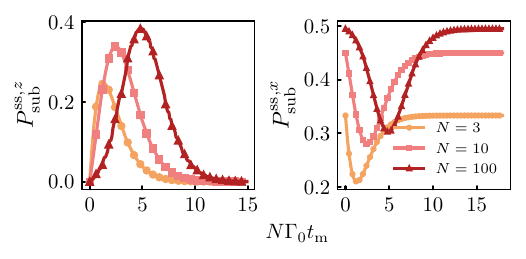}\put(40,32){\textbf{(a)}} \put(85,32){\textbf{(b)}}
\end{overpic}
\caption{Steady state probability $P_{\mathrm{sub}}^{\mathrm{ss},\mu}$ to be in a subradiant Dicke state due to a single measurement of (a) $\sigop^z$  or (b) $\sigop^x$ of a single emitter at $t_\mr{m}$ for a permutation symmetric ensemble of $N$ TLEs. Solid lines are from exact analytical expressions and symbols are from numerical solution of Eq.~\eqref{eq:SpMe}.}
\label{fig:PsubColl}
\end{figure}

\noindent\textit{Setup and Model.}--- We consider $N$ TLEs with identical energy gaps $\omega_0=ck_0=2\pi c/\lambda_0$ arranged such that they interact and emit collectively into their common electromagnetic environment. Their dynamics is described by the Gorini-Kossakowski-Sudarshan-Lindblad (GKSL) master equation for the density matrix $\hrho$ (in a frame rotating with the free hamiltonian of the TLEs $\omega_0\sum_j \sigma_j^z/2$ with $\sigop_j^{\mu = x,y,z,\pm}$ denoting the standard Pauli operators of the $j^{\mr{th}}$ TLE):
\begin{align}
   \frac{d\hrho}{dt} = \mathcal{L}_N[\hrho] \equiv -i [\Hop_\mathrm{I} , \hrho] +  \sum_{l,m=1}^N \frac{\Gamma_{lm}}{2}\Diss{\sigop_l^-}{\sigop_m^+}[\hrho] 
   \label{eq:SpMe}
\end{align}
where, $H_I = \sum_{l \neq m}\Omega_{lm}\sigop_l^+\sigop_m^-$ with $\Omega_{lm}$ the dipole-dipole interaction strength, $\Gamma_{lm}$ with $\Gamma_{mm} = \Gamma_0$ represents the correlated decay rate, and $\Diss{\hat{A}}{\hat{B}}[\cdot] = 2 \hat{A} \cdot \hat{B} - \hat{B}^\dagger \hat{A}^{\dagger} \cdot - \cdot \hat{B}^\dagger \hat{A}^{\dagger}$. 
Starting with all of the emitters in the excited state and evolving under Eq.~\eqref{eq:SpMe}, in the first protocol we interrupt this evolution by a projective measurement of the observable $\sigop_i^{z}$ or $\sigop_i^{x}$ of the $i^{\mr{th}}-$TLE at time $t_\mr{m}$. Following this measurement, we calculate steady-state occupation probability $P_{\mathrm{sub}}^{\mathrm{ss}}$ of the subradiant Dicke states $\ket{J,M,\alpha}$ (with $J<N/2$). In general, for the Dicke states we have $0\leq J \leq \frac{N}{2}$, $-J\leq M \leq J$, and degeneracy index $1\leq \alpha \leq \left\{d_J \equiv \frac{(2J+1)N!}{(\frac{N}{2}+J+1)!(\frac{N}{2}-J)!}\right\}$\footnote{Note that Dicke states with $J<\frac{N}{2}$ also have a degeneracy $d_J$, we will in general prepare a superposition over such degenerate states. See \cite{SupMat} for the exact form of the steady-state.}. In the second protocol, we perform repeated local measurements ($\sigop_i^{x}$) of the $i^\mr{th}$ TLE and track the dynamics of the population in the subradiant Dicke states of the $N-1$ unmeasured TLEs. Note that in the MCWF framework, the central quantity of interest here, $P_{\mathrm{sub}}^{\mathrm{ss}}$, is nothing but the fraction of trajectories that do not reach the ground state $\ket{J=\frac{N}{2},M=-\frac{N}{2}}$ in the long-time limit.
\begin{figure}
\centering
\begin{overpic}[width=\linewidth]{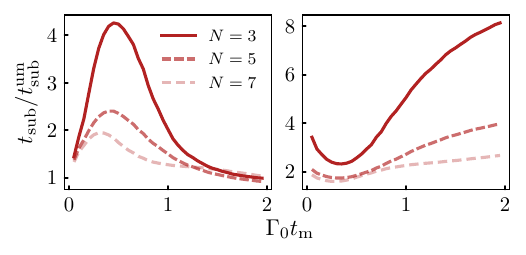}\put(40,26){\textbf{(a)}} \put(85,26){\textbf{(b)}}
\end{overpic}
\caption{Estimated life-time $t_\mr{sub}$ for an $N$ TLE array with separation $d = 0.1\lambda_0$ interacting with a waveguide as a function of $t_\mr{m}$ at which a single measurement of (a) $\sigop^z$ or (b) $\sigop^x$ is performed. Results calculated by direct numerical solution of Eq.~\eqref{eq:SpMe} are presented in units of the lifetime without measurements $t^{\mr{um}}_\mr{sub}$.}
\label{fig:t_sub}
\end{figure}

\noindent\textit{Single Measurement.}--- Let us first consider the ideal case of a PSE with $\Gamma_{lm} = \Gamma_0, \Omega_{lm} = 0, \forall l,m$. Note that in this case the collapse operator in Eq.~\eqref{eq:SpMe} simplifies to the collective form $\hat{S}^-=\sum_m \sigop_m^-$. Apart from being the prototypical scenario to study collective effects, PSEs can also be realized by symmetric configurations of TLEs coupled to waveguides \cite{Lalumiere2013}. In the absence of measurements, the MCWF trajectory unravelling of the master equation \eqref{eq:SpMe} involves a cascade down the non-degenerate maximal $J=\frac{N}{2}$ (bright) Dicke ladder $\{\ket{\frac{N}{2},M}\}$. Consider interrupting this evolution at time $t_\mr{m}$ with a measurement and say $k=\frac{N}{2}-M$ collective quantum jumps have happened. Using properties of angular momentum addition we can show that the resulting post-measurement state is given by a superposition of Dicke states in adjacent ladders \emph{i.e.} $\ket{\frac{N}{2},M},\ket{\frac{N}{2}-1,M,\alpha}$ and $\ket{\frac{N}{2},M},\ket{\frac{N}{2},M \pm 1},\ket{\frac{N}{2}-1,M,\alpha},\ket{\frac{N}{2}-1,M\pm 1,\alpha}$ for $\sigop^z$ and  $\sigop^x$ measurements respectively. As we show in detail in \cite{SupMat}, accounting for the probabilities of the post-measurement state to be in the bright Dicke subspace combined with the waiting time probability $P_{N}(k,t_\mr{m})$ to have $k$ jumps before $t_\mr{m}$ \cite{Brooke2008NearDicke,HolzingerGenes2025673679,holzinger2025solvingdickesuperradianceanalytically} allows us to write down the exact analytical expression for the subradiant population in the steady state as
\begin{align}
    P_{\mathrm{sub}}^{\mathrm{ss},\mu} = \displaystyle \sum_{k=0}^{N} P_N(k,t_\mr{m}) f^\mu(N,k), \label{eq:ExactAnalytPSE}
\end{align}
with $\mu={z,x}$ denoting the observable ($\sigop^z,\sigop^x$) being measured and $f^z(N,k) = 2k(N-k)/N^2 $, $f^x(N,k) = \frac{1}{4}+\left(\frac{(N/2-k)^2}{N^2} - \frac{1}{2N} \right)$. Eq.~\eqref{eq:ExactAnalytPSE} is one of our central results.

Fig.~\ref{fig:PsubColl} shows $P_{\mathrm{sub}}^{\mathrm{ss},\mu}(t_\mr{m})$ for a single-qubit measurement of $\sigop^z$ [(a)] and  $\sigop^x$ [(b)] calculated using Eq.~\eqref{eq:ExactAnalytPSE} as well as its clear agreement with direct numerical simulation of Eq.~\eqref{eq:SpMe} (details in \cite{SupMat}). As evident, in both cases a significant fraction of the population can be prepared in the subradiant Dicke subspace by tuning the measurement time $t_\mr{m}$. Since the measurement causes only coupling to the subradiant Dicke ladder with $J=N/2-1$, in these trajectories the system is prepared in a linear combination of the degenerate Dicke states $\{\ket{\frac{N}{2}-1,-\frac{N}{2}+1,\alpha}\}$ with a single excitation whose exact form is given in \cite{SupMat}. A key point of difference between the $\sigop^z$ and $\sigop^x$ measurement is that the former preserves the number of excitations or $M$ as opposed to the latter that can increase or decrease $M$. Thus, measuring $\sigop^z$ at very early or late times (when the state is approximately $\ket{\frac{N}{2},\pm \frac{N}{2}}$) leads to no subradiant population unlike the result for measuring $\sigop^x$. The optimal time of measurement $\Gamma_0 t_\mr{m}^\star \stackrel{N>>1}{\sim} \log N/N$ scales in the same manner as the peak radiation rate of superradiant emission and allows for a maximal population of $P_{\mathrm{sub}}^{\mathrm{ss},z}\lesssim 0.4$. In contrast, for the $\sigop^x$ measurement with large $N$, $50\%$ of the trajectories reach the dark steady-state with a single $\sigop^x$ measurement either at $t_\mr{m} \ll t_\mr{m}^\star$ or $t_\mr{m} \gg t_\mr{m}^\star$ with a minimum at $t_\mr{m}^*$. The reciprocal behavior of $P_{\mathrm{sub}}^{\mathrm{ss},z}$ and $P_{\mathrm{sub}}^{\mathrm{ss},x}$ is due to an exact relation between the two presented in \cite{SupMat} [Eq.~(S19)].

Extending our treatment beyond PSEs, we next consider an array of identical TLEs placed at separations $d$ along a 1-D waveguide with a single resonant propagating mode. This configuration results in a collective decay rate $\Gamma_{lm} = \Gamma_0\cos{(k_0 d\vert l -m \vert)} $ and dipole-dipole interaction strength $\Omega_{lm} =0.5\Gamma_0\sin{(k_0 d\vert l -m \vert)}$ \cite{Chang2012,Lalumiere2013}. Unlike the PSE, a single measurement at $t_\mr{m}$ does not lead to steady-state subradiant population here since the lack of permutation symmetry leads to the preferred unique steady state i.e. $\ket{\frac{N}{2},-\frac{N}{2}}$ with all TLEs in their ground state. Nonetheless, we find that the measurement enhances the transient population in the subradiant Dicke subspace and leads to a decay with long tail. We track this enhancement in Fig.~\ref{fig:t_sub} using the ratio of an estimate of the spontaneous life-time with measurement, $t_{\mr{sub}}$ (defined as the threshold time by which $95 \%$ if the initial excitation is emptied out), to the one without the measurement $t_{\mr{sub}}^{\mr{um}}$. The behavior has qualitative similarity to Fig.~\ref{fig:PsubColl} (b,c) in terms of the optimal time for measurement interruption to obtain large subradiant state occupation. Given our results for the PSE, a natural question that arises is whether one can obtain steady-state subradiant population for a generic collection of TLEs without permutation symmetry.

\begin{figure}
\centering
\begin{overpic}[width = 0.95\linewidth]{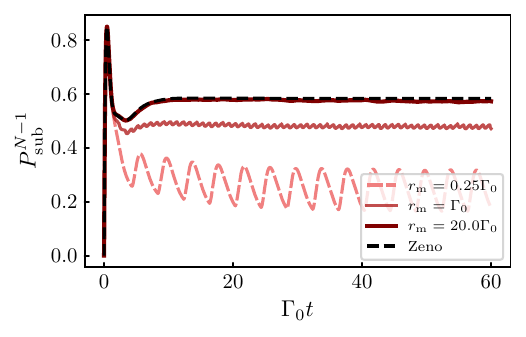}
\end{overpic}
\caption{Subradiant population of the unmeasured emitters of an array of $N=7$ TLEs placed with separation $d =0.34\lambda_0$ inside a waveguide. The $\sigop_x^i$ of the emitter at location $i=3$ is measured repeatedly at the rate $r_\mr{m}$ starting from $t_{\mr{in}} = 0.25\Gamma_0^{-1}$. Dashed lines represents the results from the Zeno limit effective master equation Eq.~\eqref{eq:ZenoME}.\label{fig:Psub_Nm1}}
\end{figure}

\textit{Multiple Measurements.}--- Towards that, in our second protocol we perform repeated measurements of $\sigop_i^x$ on the $i^{\mr{th}}$ qubit at a given rate $r_\mr{m}$ and consider the subradiant steady-state population of the unmeasured TLEs $P_{\mathrm{sub}}^{N-1}$. As Fig.~\ref{fig:Psub_Nm1} shows this indeed leads to a finite steady-state value for $P_{\mathrm{sub}}^{N-1}$ for large enough rates of measurement. Steady-state subradiant population is also confirmed by the presence of finite number of excitations in the unmeasured part as shown in \cite{SupMat}. To understand this, we derive an effective GKLS master equation in the Zeno limit \cite{Burgarth2020quantumzenodynamics,Burgarth2022oneboundtorulethem} for the density matrix projected in the Zeno subspace $\mathcal{P}_0 \hrho$ in \cite{SupMat} which in turn leads to the following evolution for the reduced density matrix of the unmeasured TLEs $\hchi = \mr{Tr}_i[\mathcal{P}_0\rhoop] = \hchi_+ + \hchi_-$ (with $\hchi_{\pm} = \prescript{}{i}{\bra{\pm_x}}\mathcal{P}_0\hrho\ket{\pm_x}_i$ and $\sigop_i^x \ket{\pm_x}_i = \pm \ket{\pm_x}_i$),
\begin{align}
    \frac{d}{dt}\hchi_{\pm} &= - i [\pm\hat{H}^\prime, \hchi_{\pm}]+ \mathcal{L}_{N-1}[\hchi_{\pm}] -\frac{\Gamma_0}{4}(\hchi_{\pm} - \hchi_{\mp}),  \label{eq:ZenoME}\\
    \Hop^\prime &=  \sum_{l \neq i} \left( \frac{\Omega_{il}} {2}\sigop_k^x\nonumber +  \frac{\Gamma_{il}}{4}\sigop_k^y \right) \nonumber.
\end{align}
Even though $\hchi$ does not evolve via a GKLS master equation, we can see that the evolution of its components in the Zeno subspace corresponding to different measurement results $\hchi_{\pm}$ is governed by an effective hamiltonian $\hat{H}^{\prime}$ that acts as an inhomogeneous local drive on the unmeasured TLEs leading to the subradiant steady state which is reminiscent of protocols using patterned driving \cite{Filipp2011,jenkins_controlled_2012,OstermannPRL2013,OstermannPRA2014,plankensteiner_selective_2015,jen_cooperative_2016,manzoni_optimization_2018,Holzinger2022} or detuning \cite{zhang_tunable_2019,parmee_signatures_2020,RubiesBigorda2022} to prepare subradiant states. As evident from solid black line in Fig.~\ref{fig:Psub_Nm1}, the Zeno limit effective master equation is very successful in reproducing the behavior from the numerics for large measurement rates. Apart from the measurement time $t_\mr{m}$ (for single measurement) and $r_\mr{m}$, we have also optimised the location of the TLE that is measured ($i$) and the array spacing $d$ to obtain the maximum $P_\mr{sub}^{N-1}$ as further discussed in \cite{SupMat}. Finally, we do not consider $\sigop^z_i$ measurement for this scheme as this leads to uncoupled evolution for the unmeasured TLEs in the Zeno limit \cite{SupMat}.
\begin{figure}
\centering
\begin{overpic}[width = 0.95 \linewidth]{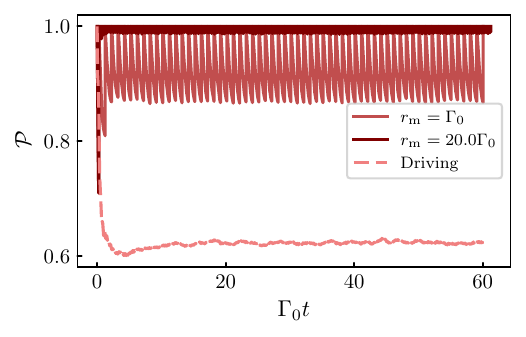}
\end{overpic}
\caption{Time evolution of the purity $\mathcal{P}$ of the state of unmeasured qubits averaged over $2000$ MCWF trajectories for different measurement rates $r_\mr{m}$ of the $i=3^\mr{rd}$ TLE. Dashed line depicts the result for driving of the $i^\mr{th}$ TLE with the Rabi frequency $\Omega_i = 10\Gamma_0$. Other parameters are same as in Fig.~\ref{fig:Psub_Nm1}.\label{fig:entanglement}} 
\end{figure}

Given that the Zeno limit can also be effectively induced by strong driving \cite{Burgarth2020quantumzenodynamics} and local driving is a well-studied strategy to prepare subradiant states \cite{Filipp2011,jenkins_controlled_2012,OstermannPRL2013,OstermannPRA2014,plankensteiner_selective_2015,jen_cooperative_2016,manzoni_optimization_2018,Holzinger2022,Fayard2023}, we finally ask how local measurements introduced here compares with the former strategy. As expected, strong driving of the $i^\mr{th}$ TLE with a coherent Rabi drive of the form $\Omega_i \sigop^x_i$ also produces subradiant Dicke population in the steady state in the undriven TLEs (see \cite{SupMat}). Remarkably, Fig.~\ref{fig:entanglement} shows that the purity of the state preparation, $\mathcal{P} = \trajavg{\mathrm{Tr}(\hrho_{N-1}^2)}$, computed from the unmeasured TLEs' reduced density matrix in individual MCWF trajectories and subsequently averaged is consistently higher in the measurement based strategy introduced here. Thus, the purifying effect of the local TLE measurements, evident in the finite measurement rate $r_\mr{m} = (20.0,1.0)$ case in  Fig.~\ref{fig:entanglement}, is a clear advantage of the scheme we have proposed. The numerical results in Fig.~\ref{fig:Psub_Nm1} are obtained by a direct numerical simulation of the master equation (with sampling over $1000$ realizations of the $\sigop^x_i$ measurements) and those in Fig.~\ref{fig:entanglement} are obtained via MCWF simulations and average over $2000$ trajectories. In both cases the QuTiP package  \cite{qutip5} was used.

\noindent \textit{Discussion}--- We have introduced two complementary protocols to generate subradiant states of a collection of TLEs based on local quantum measurement. The first scheme allows the preparation of a single excitation subradiant state whose probability can be controlled by the measurement time and its generation can also be heralded by counting the number of photons from the superradiant emission. Moreover, our preliminary analysis (see \cite{SupMat}) indicates that multiple measurements can be used to generate subradiant states with higher excitations \cite{Holzinger2022}. The second scheme exploits repeated measurements of one TLE to prepare a nearly pure state of the unmeasured TLEs with a significant overlap with the subradiant Dicke states. Moreover, the dynamics of the unmeasured TLEs can be described by the effective master equation \eqref{eq:ZenoME} in the Zeno limit which is amenable to the following interpretation - strongly measuring part of the correlated system (local control) generates effective non-uniform driving of TLEs that can be tailored by changing the $(\Gamma_{lm},\Omega_{lm})$ for instance by modifying the geometry of the array. Furthermore, in both schemes we find that the steady state prepared, as expected for subradiant states, has finite entanglement \cite{SupMat}.

Coming to the experimental feasibility of the proposals, note that our schemes are platform agnostic and can be implemented in various architectures such as superconducting qubits \cite{Lalumiere2013,vanLoo2013PhotonMediatedArtificialAtoms,Mlynek2014TwoQubitSuperradiance,Mirhosseini2018SCMetamaterialsWQED,Wang2020SCQubitSubradiance,Kannan2021WaveguideBandgapSCArray,Zanner2022DarkStateWQED,Sheremet2023WQEDReview,Sharafiev2025CollectiveEffectsSC}, trapped ions \cite{DeVoe1996SuperSubradiantTwoIons} and neutral atoms \cite{Glicenstein2020,Rui2020SubradiantOpticalMirror, Yan2023SuperradiantSubradiantCavity,Hofer2025}, quantum dots \cite{Scheibner2007QDSuperradiance,Kim2018QDWaveguideSuperradiance,Tiranov2023AtomicArraySubradiance,kim_cavity-mediated_2025}, NV centers \cite{Bradac2017NanodiamondSuperradiance,Angerer2018SuperradiantColourCentres,qu2024superradiancenitrogenvacancycenters}, and ultracold atoms in free space \cite{Inouye1999SuperradiantRayleighBEC,Bienaime2010CooperativeScattering,
Bachelard2012SubradiantModes,Pellegrino2014SuppressionScattering,Guerin2016ColdAtomSubradiance,Jennewein2016CoherentScattering,Jenkins2016CollectiveFluorescence,Chen2018OneDSuperradianceLattice,Ferioli2021,Ferioli2023NonEquilibriumSuperradiant,Agarwal2024DirectionalSuperradiance} as well as coupled to cavities \cite{Slama2007SuperradiantRayleighPRA,Bohnet2012SteadyStateSuperradiantNature,Norcia2016SuperradiantStrontiumSciAdv,Bohr2024CollectivelyEnhancedRamseyNatComm,Gabor2025SubradiantAtomArrayEPJQT,BaumgartnerDonner2025} or waveguides \cite{Chang2014SuperradianceAtomsPCWaveguide,
Goban2015SuperradiancePCW,Hood2016AtomAtomBandEdgePCW,Solano2017SuperradianceInfiniteRangeNC,Pennetta2022CollectiveRadiativeNanofiber,Liedl2024SuperradiantBurstsNanofiber,Zhou2024} where collective dissipative phenomena have been successfully demonstrated. The only key requirement for the implementation is the ability to perform measurements on individual TLEs over a time scale $T_\mr{meas}$ significantly shorter than the superradiant decay time-scale which is $\lesssim (N\Gamma_0)^{-1}$. While the technology to support this is available across platforms, the ideal setting would be superconducting qubits where high fidelity fast measurement gates with $T_\mr{meas}$ several orders of magnitude smaller than the dissipation time-scales have been implemented \cite{Walter2017FastReadout,Opremcak2018OnChipPhotonCounter,Sunada2022PurcellFilter,Quarton2024UltrafastSCReadout}.

\noindent \textit{Conclusions}---Subradiant states are naturally occupied during the spontaneous emission dynamics of sub-wavelength arrays that do not have permutation symmetry and manifest as long-tails in the ring-down to the ground state. This indicates that while there is a coupling between bright and dark Dicke state subspaces during such an evolution, this a two-way coupling with the eventual steady state given by the trivial Dicke `bright' state with all the emitters in their ground state. In contrast, for a permutation symmetric ensemble there is no coupling between the bright and dark Dicke subspaces. We have shown that local quantum measurements are able to controllably and irreversibly couple the bright and dark subspaces in PSEs leading to steady state occupation of subradiant states. In non PSEs, the Zeno effect engendered by strong quantum measurement on one of the TLEs of an ensemble evolving in a correlated manner can be leveraged to direct the unmeasured TLEs to subradiant steady states. Our study opens up several directions for future research such as the generalization to multiple measurements in the PSE case as well as a detailed study of entanglement dynamics \cite{ZhangRabl2025} especially in the light of the highly topical results extending measurement induced phase transitions to open quantum systems \cite{EhudAltman,Biella2021manybodyquantumzeno}.

\begin{acknowledgments}
We thank Athreya Shankar for useful discussions. I.B. and B.P.V. acknowledge funding from the DST National Quantum Mission through the project DST/FFT/NQM/QSM/2024/3. I.B. acknowledges support via a CSIR-UGC NET Ph.D. fellowship. A.T. is supported by a Sabarmati
Fellowship of IIT Gandhinagar.  B.P.V acknowledges support from MATRICS Grant No. MTR/2023/000900 from Anusandhan National Research Foundation, Government
of India.
\end{acknowledgments}
\bibliography{references.bib}
\end{document}


\title{Supplemental Material: Measurement Induced Subradiance}
\author{Ipsita Bar}
\affiliation{Indian Institute of Technology Gandhinagar, Palaj, Gujarat 382355, India}
\author{Aditi Thakar }
\affiliation{Indian Institute of Technology Gandhinagar, Palaj, Gujarat 382355, India}
\author {B. Prasanna Venkatesh}
\affiliation{Indian Institute of Technology Gandhinagar, Palaj, Gujarat 382355, India}
\maketitle

\section{Additional Details Regarding Permutation Symmetric Ensembles}
\subsection{Steady-state subradiant population calculation}
\begin{figure}
\centering
\begin{overpic}[width = 0.95 \linewidth]{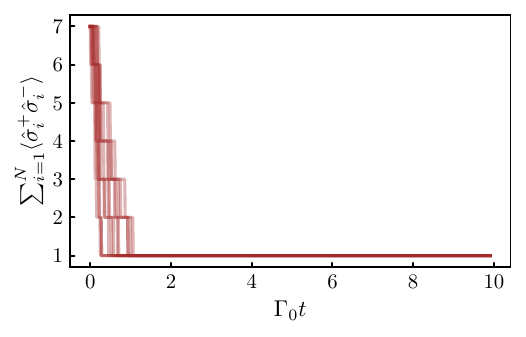}
\end{overpic}
\caption{Individual MCWF trajectories freezing to the single excitation dark state of the Dicke ladder $J= \frac{N}{2} -1$ with $N=7$ for the permutationally symmetric ensemble when a single measurement of $\sigop^z_i$ is performed at time $\Gamma_0 t_\mr{m} = 0.25$. }
\label{fig:supp1}
\end{figure}
In this section, we detail the calculation of the steady-state probability of producing a subradiant state in a permutation symmetric ensemble (PSE) of $N$ TLEs decaying from the excited state $\ket{E} = \ket{J=N/2,M=N/2} = \bigotimes_{i=1}^{N}\ket{e_{i}}$ by interrupting its evolution with a single measurement of $\sigop^{z/x}_i$ at time $t_\mr{m}$. Note that, $\ket{e_i}$ ($\ket{g_i}$) denotes the ket with $i^\mr{th}$ TLE in the excited (ground) state and $\sigop_i^z \ket{e_i} = + \ket{e_i}$ ($\sigop_i^z \ket{g_i} = - \ket{g_i}$). Let us consider first a Monte-Carlo wavefunction (MCWF) quantum trajectory unraveling \cite{Dalibard1992MCWF,Carmichael1993OpenSystems,Molmer1993MCWF,Molmer1996ReviewMCWF} of the GKLS master equation for a PSE:
\begin{align}
   \frac{d\hrho}{dt} = \frac{\Gamma_0}{2} \left[2 \Sop^- \hrho \Sop^+ - \Sop^+ \Sop^- \hrho - \hrho \Sop^+ \Sop^-\right], 
   \label{eq:SpMe}
\end{align}
with the only jump operator for evolution given by the collective one $\Sop^{\pm} = \sum_{i=1}^{N} \sigop^\pm_i$. The MCWF trajectory involves non-hermitian evolution of the state $\ket{\psi(t)}$ from $t$ to $t+\Delta t$ with the hamiltonian (setting $\hbar = 1$ for notational simplicity) $\Hop_\mr{nh} = - i \frac{\Gamma_0}{2}\Sop^+\Sop^-$ interrupted by quantum jumps with a probability $\delta p = \Gamma_0\Delta t \sandwich{\psi(t+\Delta t)}{\Sop^+\Sop^-}{\psi(t+\Delta t)}$. In the absence of any measurements, this leads to a superradiant cascade down the bright or maximal $J$ Dicke ladder to the ground state $\ket{G} = \ket{J=N/2,M=-N/2} = \bigotimes_{i=1}^{N}\ket{g_{i}}$ with $N$ jumps over an average time $\Gamma_0\avg{\tau_{sr}} \sim \log(N)/N$. 

Let us now consider a single measurement of $\sigop^{\mu}_i$ ($\mu=x,z$) of one of the TLEs at time $t_\mr{m}$. The population of the subradiant Dicke manifold (states $\ket{J<N/2,-J,\alpha}$) in the steady state under such measured evolution is essentially the fraction of MCWF trajectories that do not end up in the bright ground state $\ket{G}$ but freeze to a finite excitation dark Dicke state as shown in Fig.~\ref{fig:supp1}. Given all the trajectories generated by the MCWF method contribute equally to the averaged steady state $\hrho^\mr{ss}$, the fraction of trajectories that terminate in the ground state is the probability $P_{G}^{\mr{ss},\mu}(t_\mr{m})$ that a trajectory in the presence of a measurement at $t_\mr{m}$ ends up in $\ket{G}$. Thus, our aim is to calculate $P_{G}^{\mr{ss},\mu}(t_\mr{m})$. Towards that we write:
\begin{align}
    P_{G}^{\mr{ss},\mu}(t_\mr{m}) = \displaystyle \sum_{k=0}^{N} P_N(k,t_\mr{m}) g^\mu(N,k) \label{eq:ExactAnalytPSE},
\end{align}
with $\mu={z,x}$ denoting the observable ($\sigop^z,\sigop^x$) being measured. Here $P_N(k,t_\mr{m})$ is the probability that there have been $k$ quantum jumps in the MCWF trajectory before the measurement at $t_\mr{m}$ and the function $g^\mu(N,k)$ is the probability that the post-measurement state terminates in $\ket{G}$. 

The quantity $P_N(k,t_\mr{m})$ is the waiting time distribution for superradiant emission \emph{i.e.} the probability that a PSE has emitted $k$ photons by the time $t_\mr{m}$. This has been calculated analytically previously \cite{Brooke2008NearDicke,holzinger2025solvingdickesuperradianceanalytically}. For the sake of completeness, we present the result in the form we use in our analytical computation. The two major ways to define $P_N(k,t_\mr{m})$ are via the following recursion relation
\begin{align}
  P_N(k,t_\mr{m}) = \lambda_{k-1}\int_0^{t_{\mr{m}}} d \tau e^{\lambda_k (\tau-t_m)}P_N(k-1,t_\mr{m}), \label{eq:recursivePN}
\end{align}
with $\lambda_k = (N-k)(k+1)$ and $P_{N}(0,t_\mr{m}) = e^{-Nt_\mr{m}}$ or the associated differential equation
\begin{align}
  \frac{dP_N(k,t)}{dt} = -\lambda_{k}P_N(k,t) + \lambda_{k-1} P_N(k-1,t) \label{eq:PNdiffeq},  
\end{align}
with initial condition $P_N(k\geq0,t=0) = 0$. The differential equation admits a closed form solution in Laplace space and can be expressed as
\begin{align}
    P_N(k,t_\mr{m}) = \mathscr{L}^{-1}\left\{ \frac{\prod_{j=0}^{k-1} \lambda_j }{\prod_{j=0}^k(s+\lambda_j)}\right\}(t_\mr{m})\label{eq:invLaplace},
\end{align}
with $\mathscr{L}^{-1}$ denoting the inverse Laplace transformation (ILT). Moreover, the ILT can be performed explicitly leading to the solution
\begin{align}
    &P_N(k,t_\mr{m}) =  \frac{k!N!}{(N-k)!} \left[ 
    \sum_{\lambda_r \in \mr{non-deg}}\prod_{j\neq r}^k \frac{e^{-\lambda_r t_\mr{m}}}{\lambda_j-\lambda_r} \right . \nonumber \\ 
    & \left.+ \sum_{\lambda_r \in \mr{deg}} (A_r+B_r t_\mr{m})e^{-\lambda_r t_m}  \label{eq:PNexact}\right] \\
   & B_r = \prod_{j \neq r,(N-1-r)}^k \frac{1}{\lambda_j-\lambda_r}, A_r =  \sum_{j \neq r,(N-1-r)}^k \frac{-B_r}{\lambda_j-\lambda_r}\nonumber.
\end{align}
Note that in the above solution we have explicitly separated the parts where $\lambda_r$ are degenerate which happens when $k>N/2$.
\begin{figure}
\centering
\begin{overpic}[width = 0.95 \linewidth]{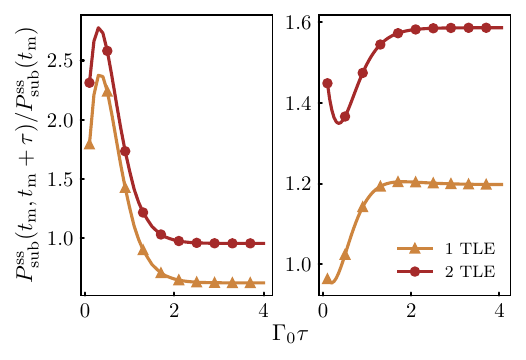}\put(35,50){\textbf{(a)} $\sigop^z$}\put(80,50){\textbf{(b)} $\sigop^x$}
\end{overpic}
\caption{Steady state subradiant population for $\sigop^z$ (a) $\sigop^x$ (b) measurements at times $\{t_\mr{m},t_\mr{m}+\tau\}$ on either the same (triangles) or two different (filled circles) emitters of an $N=4$ TLE permutation symmetric ensemble. $\Gamma_0 t_{\mr{m}} = 0.005$ and the population is expressed in units of the subradiant population with a single measurement.}
\label{fig:supp2}
\end{figure}

We calculate the probability that the post-measurement state terminates in $\ket{G}$, $g^\mu(N,k)$, as follows. The state before the measurement in the case with $k$ jumps before $t_\mr{m}$ is $\ket{\psi_k} = \ket{N/2,N/2-k}$. Formally, we can write
\begin{align*}
    g^\mu(N,k) = p_{\sigma^\mu_i=+|k} p_{\mu,\mr{bright},+}+ p_{\sigma^\mu_i=-|k} p_{\mu,\mr{bright},-},
\end{align*}
where $p_{\sigma^\mu_i=\pm|k}$ denotes the probability that a measurement of $\sigop^\mu_i$ results in the outcome $\pm 1$ given the state $\ket{\psi_k}$ and $p_{\mu,\mr{bright},\pm}$ is the probability that the associated post-measurement state is in the bright Dicke subspace $\ket{J=N/2,M}$. The key property that enables this calculation is the decomposition of $\ket{\psi_k}$ using angular momentum algebra as
\begin{figure}
\centering
\begin{overpic}[width = 0.95 \linewidth]{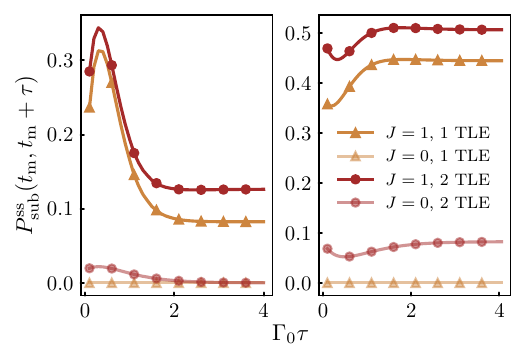}\put(35,50){\textbf{(a)} $\sigop^z$}\put(80,50){\textbf{(b)} $\sigop^x$}
\end{overpic}
\caption{Steady state subradiant population resolved into different Dicke ladders with angular momentum values $J$ for $\sigop^z$ (a) $\sigop^x$ (b) measurements at times $\{t_\mr{m},t_\mr{m}+\tau\}$ on either the same (triangles) or two different (filled circles) emitters. Other parameters same as in Fig.~\ref{fig:supp2}}
\label{fig:supp3}
\end{figure}
\begin{align}
    \ket{\psi_k} &= \sqrt{\frac{N-k}{N}} \left \vert \frac{1}{2},\frac{1}{2} \right \rangle_i \otimes \left \vert\frac{N-1}{2},k-\frac{1}{2} \right \rangle_{N-1} \nonumber \\
    &+\sqrt{\frac{k}{N}} \left \vert \frac{1}{2},-\frac{1}{2} \right \rangle_i \otimes \left \vert\frac{N-1}{2},k+\frac{1}{2} \right \rangle_{N-1} \label{eq:CGdecomp},
\end{align}
where we have used a modified notation for the eigenkets of the $i^\mr{th}$ TLE $\ket{\frac{1}{2},\frac{1}{2}}_i = \ket{e}_i$ and $\ket{\frac{1}{2},-\frac{1}{2}}_i = \ket{g}_i$ and $\ket{\frac{N-1}{2},M}_{N-1}$ are the bright Dicke states of $N-1$ unmeasured TLEs. Using the above state we can immediately calculate (with $m_k = N/2-k$)
\begin{align}
p_{\sigma^z_i=\pm|k} =  \frac{\frac{N}{2}\pm m_k}{N} \label{eq:sigzxprobmeas}, \,\,
p_{\sigma^x_i=\pm|k} =  \frac{1}{2}.
\end{align}

Let us next determine $p_{z,\mr{bright},\pm}$ by noting first that the post-measurement state for the two outcomes of the $\sigop^z_i$ measurement can be written as
\begin{align}
&\ket{\psi_{z,\pm}} = \ket{\frac{1}{2},\pm \frac{1}{2}} \otimes \ket{\frac{N-1}{2},k\mp \frac{1}{2}}_{N-1}, \label{eq:postmeasz}\\
&=c_{N/2,k\vert\pm} \ket{\frac{N}{2},k}_N + \sum_{\alpha=1}^{d_{N/2-1}} c_{N/2-1,k,\alpha\vert \pm}  \ket{\frac{N}{2}-1,k,\alpha}_N \label{eq:postmeaszcoll},
\end{align}
where the second line follows from properties of angular momentum addition and for added clarity we have used the subscript $N$ ($N-1$) on the Dicke states to denote the corresponding number of TLEs.
Eq.~\eqref{eq:postmeaszcoll} clearly demonstrates how the measurement causes a population redistribution between the maximal (bright) Dicke ladder with $J=N/2$ and the (degenerate) subradiant Dicke ladder with $J=N/2-1$. Subsequent time-evolution for a PSE proceeds in an independent manner in the two ladders with the bright ladder terminating in the trivial ground state $\ket{G}$ and the subradiant ladder in the dark state subspace $\ket{\frac{N}{2}-1,-(\frac{N}{2}-1),\alpha}$. Using Eq.~\eqref{eq:postmeaszcoll} and ~\eqref{eq:postmeasz} we can calculate the coefficients $c_{N/2,k\vert\pm}$ and immediately write the probability to be in the bright state ladder as
\begin{align}
    p_{z,\mr{bright},\pm} = \vert c_{N/2,k\vert\pm} \vert^2 = \frac{\frac{N}{2} \pm m_k}{N}.\label{eq:pbrightz}
\end{align}
\begin{figure}
\centering
\begin{overpic}[width = 0.95 \linewidth]{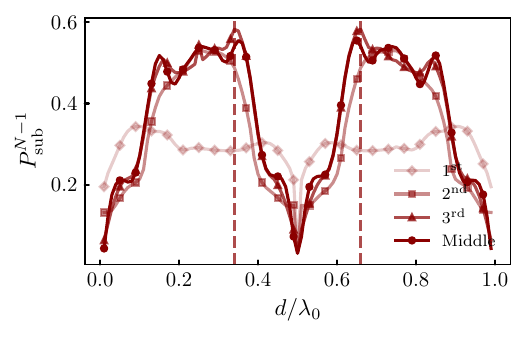}
\end{overpic}
\caption{Steady-state population in the Zeno-limit [Eq.~\eqref{eq:chievolmain}] of the unmeasured $N-1$ TLEs coupled to a waveguide as a function of their separation $d/\lambda_0$ for $N = 7$. Maximum $P_{\mr{sub}}^{N-1}$ is obtained at $d = 0.34\lambda_0$ and $0.66\lambda_0$ (dashed lines) for measurement on the $3^{\mr{rd}}$ qubit.}
\label{fig:supp4}
\end{figure}

A similar analysis can be repeated for the $\sigop^x_i$ measurement with the difference that the post-measurement state takes the form
\begin{align}
    \ket{\psi_{x,\pm}} &= \sqrt{\frac{N-k}{N}} \ket{\pm_x}_i \otimes \left \vert\frac{N-1}{2},k-\frac{1}{2} \right \rangle_{N-1} \nonumber \\
    &\pm\sqrt{\frac{k}{N}} \ket{\pm_x}_i \otimes \left \vert\frac{N-1}{2},k+\frac{1}{2} \right\rangle_{N-1} \label{eq:postmeasx},
\end{align}
with $\ket{\pm_x}_i = \frac{1}{\sqrt{2}}(\ket{e}_i\pm \ket{g}_i)$ are $\sigop^x_i$ eigenkets. Moreover, the post-measurement state can also be written in terms of collective states as:
\begin{align}
 \ket{\psi_{x,\pm}} &= \sum_{j=0,\pm 1}\tilde{c}_{N/2,k + j\vert\pm} \ket{\frac{N}{2},k+j}_N \label{eq:postmeasxcoll}\\ 
 &+ \sum_{\alpha=1}^{d_{N/2-1}} \tilde{c}_{N/2-1,k,\alpha\vert \pm}  \ket{\frac{N}{2}-1,k,\alpha}_N \nonumber.  
\end{align}
Note that $\sigop^x_i$ measurement also causes coupling to adjacent states of different $M$ (excitations) within the Dicke ladder in addition to the coupling with adjacent subradiant Dicke ladder with $J=N/2-1$. In this sense the $\sigop^x_i$ measurement can pump or drain excitations from the PSE. Using Eqs.~\eqref{eq:postmeasx} and ~\eqref{eq:postmeasxcoll} we get
\begin{align}
    p_{x,\mr{bright},\pm} &=  \sum_{j=0,\pm 1} \vert \tilde{c}_{N/2,k + j\vert\pm} \vert^2 = \frac{3}{4}-\left( \frac{m_k^2}{N^2} - \frac{1}{2N} \right)
    \label{eq:pbrightx}
\end{align}
Combining Eqs.~\eqref{eq:sigzxprobmeas},\eqref{eq:pbrightz}, and ~\eqref{eq:pbrightx}, we have finally
\begin{align}
g^z(N,k) &=  1- \frac{2k(N-k)}{N^2}, \label{eq:gzNk} \\
g^x(N,k) &= \frac{3}{4}-\left( \frac{\left(\frac{N}{2}-k \right)^2}{N^2} - \frac{1}{2N} \right)\label{eq:gxNk}.
\end{align}
The probability of reaching a subradiant steady state given a measurement after $k$ collective spontaneous emission steps $f^{\mu}(N,k)$ quoted in the main paper are then easily derived as $f^{\mu}(N,k) = 1-g^{\mu}(N,k)$ leading to the main analytic result of our work
\begin{align}
    P_\mr{sub}^{\mr{ss},z} &= \sum_{k=0}^N P_N(k,t_\mr{m}) \frac{2k(N-k)}{N^2} \label{eq:Psubssz},\\
    P_\mr{sub}^{\mr{ss},x} &= \sum_{k=0}^N P_N(k,t_\mr{m}) \left[ \frac{1}{4} + \frac{\left(\frac{N}{2}-k \right)^2}{N^2} - \frac{1}{2N} \right]. \label{eq:Psubssx}
\end{align}
\begin{figure}
\centering
\begin{overpic}[width = 0.95 \linewidth]{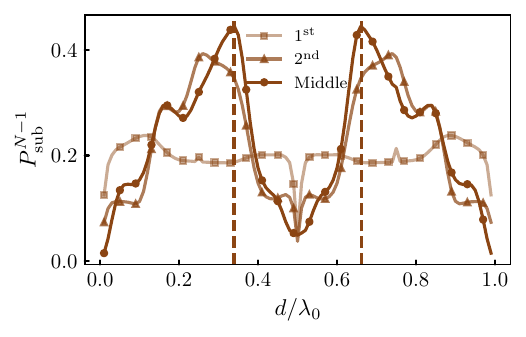}
\end{overpic}
\caption{Steady-state population in the Zeno-limit [Eq.~\eqref{eq:chievolmain}] of the unmeasured $N-1$ TLEs coupled to a waveguide as a function of their separation $d/\lambda_0$ for $N = 5$. Maximum $P_{\mr{sub}}^{N-1}$ is obtained at $d = 0.34\lambda_0$ and $0.66\lambda_0$ (dashed lines) for measurement on the $3^{\mr{rd}}$ qubit.}
\label{fig:supp5}
\end{figure}
Interestingly, from the above expressions, noting that $P_N(k,t_\mr{m})$ is a normalized probability distribution, we obtain the relation
\begin{align}
    P_\mr{sub}^{\mr{ss},x} = \frac{N-1}{2N}-\frac{P_\mr{sub}^{\mr{ss},z}}{2}, \label{eq:Psubxsrelation}
\end{align}
between the subradiant steady-state populations for the $\sigop^x$ and $\sigop^y$ measurements that captures their general reciprocal mirror-like behavior shown in Fig.~2 of the main paper.

The analytical expressions Eqs.~\eqref{eq:Psubssz} and \eqref{eq:Psubssz} essentially allow the calculation of the subradiant population for any $N$. While the closed form expressions for small $N=2,3$ such as
\begin{figure}
\centering
\begin{overpic}[width = 0.95 \linewidth]{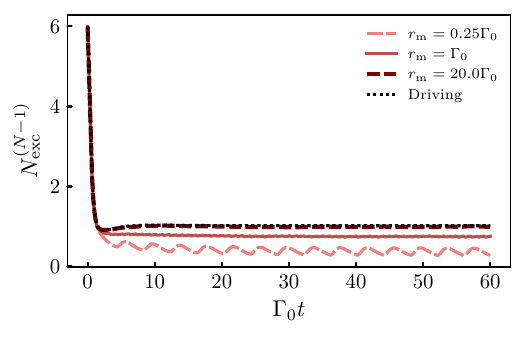}
\end{overpic}
\caption{Time evolution of the total number of excitations in the unmeasured $N-1$ TLEs of an array of $N=7$ emitters with separation $d = 0.34\lambda_0$. The $i=3^{\rm{rd}}$ emitter is measured at different rates $r_\mr{m}$ (dash dotted, solid, dashed lines) or driven strongly with Rabi frequency $\Omega_i =10\Gamma_0$ (dotted black line).}
\label{fig:supp6}
\end{figure}
\begin{align*}
    P_\mr{sub}^{\mr{ss},z}(N=2) &= t_\mr{m}e^{-2\Gamma_0 t_\mr{m}},\\
    P_\mr{sub}^{\mr{ss},z}(N=3) &= \frac{4}{3}e^{-4\Gamma_0 t_\mr{m}} \left[3+ e^{\Gamma_0 t_\mr{m}}(-3+4\Gamma_0 t_\mr{m}) \right],
\end{align*}
are tractable, for larger $N$ they are sums of many exponential factors. Thus, let us briefly remark on some general properties of the resulting $P_\mr{sub}^{\mr{ss},z/x}$. $P_\mr{sub}^{\mr{ss},z}(t_\mr{m})$ goes to $0$ as $t_\mr{m}\rightarrow 0$ (as a power-law in $\Gamma_0 t_\mr{m}$) and $t_\mr{m} \rightarrow \infty$ (as an exponential $e^{-N \Gamma_0 t_\mr{m}}$), with a peak value at $t_\mr{m}^\star \sim \log(N)/(N \Gamma_0)$. The asymptotic approach to zero at small and large $t_\mr{m}$ is easily understood from the fact that the PSE is either in the ground or excited state in these limits and a $\sigop^z$ measurement leaves the state unchanged. At an intermediate time $t_\mr{m}^* \sim \log(N)/(N \Gamma_0)$ coincidental with the time for the radiated power's peak for superradiant emission, $P_\mr{sub}^{\mr{ss},z}$ takes its maximum value. This behavior can also be understood from Eqs.~\eqref{eq:pbrightz} and \eqref{eq:gzNk} which show that the probability to collapse to a bright state upon measurement takes its minimum value when $k \sim N/2$ jumps have taken place before the measurement. Consequently, for the maximum transfer to subradiant Dicke ladder, a $\sigop^z$ measurement at the time when the system is most likely to occupy the state $m_k \sim 0$ at the middle of the bright Dicke ladder which is coincident with the radiation emission peak is most effective. In contrast, as evident from Eq.~\eqref{eq:Psubxsrelation}, the behavior of $P_\mr{sub}^{\mr{ss},x}(t_\mr{m})$ is complementary with a minimum value at $t_\mr{m}^* \sim \log(N)/(N \Gamma_0)$ and maximum taken at $t_\mr{m} \rightarrow {0,\infty}$. The minimum can again be understood by appealing to Eq.~\eqref{eq:pbrightx} showing that the probability to remain in the bright subspace post-measurement is maximum when $m_k \sim 0$ \emph{i.e.} at the middle of the Dicke ladder. In contrast, using Eq.~\ref{eq:Psubxsrelation} and $P_\mr{sub}^{\mr{ss},z}(t_\mr{m} \rightarrow \{0,\infty \}) = 0$, it is immediately clear that
\begin{align}
    P_\mr{sub}^{\mr{ss},x}(t_\mr{m} \rightarrow \{0,\infty \}) = \frac{1}{2} - \frac{1}{2N} \label{eq:zerotimePssx},
\end{align}
underscoring our finding that even a single $\sigop^x$ measurement from the fully excited or ground state leads to a single-excitation subradiant Dicke state with a probability of $\sim 1/2$ for large $N$.
\begin{figure}
\begin{centering}
\subfloat{\begin{overpic}[width = 0.95 \linewidth]{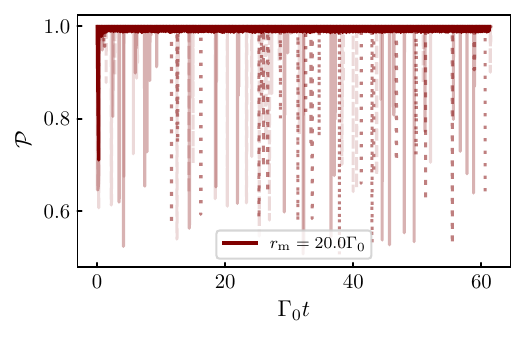}
\put(25,27){\textbf{(a)}}
\end{overpic}
}\\
\subfloat{\begin{overpic}[width = 0.95 \linewidth]{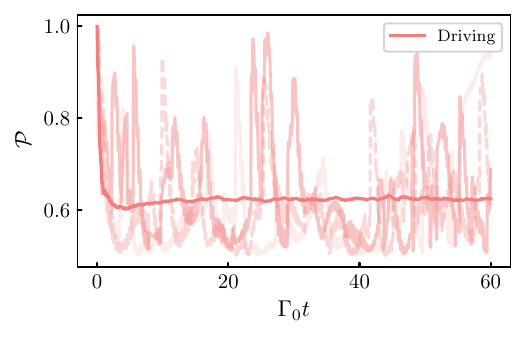}
\put(25,27){\textbf{(b)}}
\end{overpic}
}
\end{centering}
\caption{Purity of the unmeasured TLEs' state for multiple measurements performed on the third qubit of an array of TLEs with $N=7$ and $d = 0.34\lambda_0$. Here, $\hrho_{N-1}^2$ is calculated over each MCWF trajectory (translucent lines) and averaged over $N_{\mr{traj}} = 2000$ trajectories for measurement rate (a) $r_{\mr{m}} = 20.0\Gamma_0$ and (b) driving the $3^{\mr{rd}}$ qubit with Rabi frequency $\Omega_i =10 \Gamma_0$.} 
\label{fig:supp7}
\end{figure}
\subsection{Entanglement entropy in the dark steady state}
The finite subradiant Dicke state probability in the steady state also implies that the steady-state we have prepared using the local measurement has finite entanglement. Let us analyse the same by writing down the form of the subradiant state prepared. As discussed above in Eqs.~\eqref{eq:postmeasz} and ~\eqref{eq:postmeasx}, the post-measurement state after the measurement at time $t_\mr{m}$ can be written within the $2N$ dimensional subspace of the measured emitter and the collective bright states of unmeasured emitters. The general state of the TLEs in the MCWF trajectory that follows can then be expanded as:
\begin{align}
    \ket{\psi(t)} = \sum_{M=-\frac{N-1}{2}}^{\frac{N-1}{2}} \sum_{m = \pm \frac{1}{2}} \ket{\frac{1}{2},m} \otimes \ket{J=\frac{N-1}{2},M}_{N-1} \label{eq:genstate}.
\end{align}
The exact form of the subradiant Dicke state can be worked out by requiring $\Sop_N^- \ket{\psi(t)} = 0$ and ensuring that the trivial ground state $\ket{G}_N = \ket{\frac{1}{2},-\frac{1}{2}} \otimes \ket{J=\frac{N-1}{2},-\frac{(N-1)}{2}}_{N-1}$ is excluded from the sum. This leads to the following form for the subradiant dark state
\begin{align}
\ket{\psi_D} &= \frac{1}{\sqrt{2}}\ket{\frac{1}{2},\frac{1}{2}} \otimes \ket{J=\frac{N-1}{2},-\frac{(N-1)}{2}}_{N-1} \label{eq:darkstate}\\
& - \frac{1}{\sqrt{2}} \ket{\frac{1}{2},-\frac{1}{2}} \otimes \ket{J=\frac{N-1}{2},-\frac{(N-1)}{2}+1}_{N-1}\nonumber.
\end{align}
From angular momentum addition, it is immediately clear that this state is in the degenerate subradiant Dicke subspace  $\{\ket{J=N/2-1,M=-(N/2-1),\alpha}_N\}$ of $N$ TLEs. Since the entanglement entropy is a non-linear function of the state, the correct way to compute it for collective spontaneous emission \cite{delmonte_measurement-induced_2025} is to first calculate it for single MCWF trajectories and subsequently average over many trajectories. In our case, given that the two possible steady-states (upon a single measurement at $t_\mr{m}$) for the MCWF trajectories is to terminate in the ground state $\ket{G}_N$ with probability $P_G^{\mr{ss},\mu}$ or in the dark state $\ket{\psi_D}$ with probability $P_\mr{sub}^{\mr{ss},\mu}$, the average entanglement entropy is
\begin{align}
    S^\mr{ss,\mu} = P_\mr{sub}^{\mr{ss},\mu} S_{N/2}\left [\ket{\psi_D}\bra{\psi_D} \right]. \label{eq:entropyformal}
\end{align}
Here $S_{N/2}[\ket{\psi_D}\bra{\psi_D}]$ represents the half-chain entropy of the subradiant state and can be computed as follows. Let us partition the $N$ TLE system into two parts $A$ and $B$ with $N_A$ and $N_B$ spins respectively. Assuming that the measured qubit (which is assigned the index $1$ without loss of generality) is in the partition $A$, we can write
\begin{align*}
    \ket{\psi_D} &= \ket{\phi_0}_A \otimes \ket{G}_B + \ket{\phi_1}_A \otimes \ket{W}_B, \\
    \ket{\phi_0}_A &= \frac{\ket{e}_1 \otimes \ket{G}_{N_A-1} - \sqrt{\frac{N_A-1}{N-1}}\ket{g}_1 \otimes \ket{W}_{N_A-1}}{\sqrt{2}}, \\
    \ket{G}_A &=\bigotimes_{j=2}^{N_A}\ket{g}_j,\,  \ket{G}_B = \bigotimes_{j=N_A+1}^{N} \ket{g}_j,\\
    \ket{\phi_1}_A &= -\sqrt{\frac{N_B}{2(N-1)}} \ket{g}_1 \otimes \ket{G}_A,\\
    \ket{W}_A &= \frac{1}{\sqrt{N_A-1}} \sum_{j=2}^{N_A} \ket{e}_j \otimes \bigotimes_{k \neq j}^{N_A}\ket{g}_k,\\
    \ket{W}_B &= \frac{1}{\sqrt{N_B}}\sum_{j=1}^{N_B} \ket{e}_j \otimes \bigotimes_{k \neq j}^{N_B}\ket{g}_k.
\end{align*}
From the fact that $\ket{G}_B$ and $\ket{W}_B$ are orthonormal, the reduced density matrix over $B$ is given by
\begin{align*}
    \hrho_A = \mr{Tr}_{B} [\ket{\psi_D}\bra{\psi_D}]
    = \ket{\phi_0}_A\bra{\phi_0} + \ket{\phi_1}_A\bra{\phi_1}.
\end{align*}
The entanglement entropy can then by written as the von Neumann entropy of $\hrho_A$ and is given by
\begin{align*}
    S_A &= -\frac{1}{2} \left(1+\frac{N_A-1}{N-1}\right) \log \left [ \frac{1}{2} \left(1+\frac{N_A-1}{N-1}\right) \right] \\
    &-\frac{N_B}{2(N-1)} \log \left (\frac{N_B}{2(N-1)} \right)
\end{align*}
Taking $N_A = N_B = N/2$, we get the final answer for the entanglement entropy in the steady state as
\begin{align}
    S^\mr{ss,\mu} &= P_\mr{sub}^{\mr{ss},\mu} \left[ -\frac{3N-4}{4(N-1)} \log \left(\frac{3N-4}{4(N-1)} \right)  \right . \nonumber\\
    & \left . -\frac{N}{4(N-1)} \log \left (\frac{N}{4(N-1)} \right)\right ]\label{eq:entropyexact},
\end{align}
which takes the following ($N$-independent) form for large $N$
\begin{align*}
    S^\mr{ss,\mu} \stackrel{N \rightarrow \infty}{\sim} P_\mr{sub}^{\mr{ss},\mu}\left(2 \log 2- \frac{3}{4} \log 3 \right).
\end{align*}

\subsection{Multiple Measurements}

Let us next comment on the possibilities that multiple measurements on a PSE opens up. We first note that additional $\sigop^{z,x}$ measurements on the \emph{same} TLE can also only lead to subradiant Dicke states with single excitation. This follows from the decompositions Eqs. ~\eqref{eq:postmeasz} and ~\eqref{eq:postmeasx} of the PSE state after measurement on the $i^\mr{th}$ TLE. Since further measurements on the $i^{th}$ TLE only affect the state of the measured TLE, a similar decomposition and the associated expansions in terms of collective $N-$ TLE states as in Eqs.~\eqref{eq:postmeaszcoll} and ~\eqref{eq:postmeasxcoll} will continue to hold after such measurements. Thus, the population will be restricted to just the two Dicke ladders with $J=N/2,N/2-1$. Denoting the population in the subradiant states due to measurements at times $\{t_\mr{m},t_\mr{m}+\tau\}$ denoted by $P_\mr{sub}^{\mr{ss},\mu}(t_\mr{m},t_{\mr{m}}+\tau)$, as shown in Fig.~\ref{fig:supp2}, it is clear that additional measurements can enhance the steady-state subradiant population with respect to the one obtained from a single measurement at $t_\mr{m}$ \emph{i.e.} $P_\mr{sub}^{\mr{ss},\mu}(t_\mr{m})$. Interestingly, performing two measurements on two distinct TLEs leads to an even higher enhancement. Moreover, this population is distributed in a higher excitation sub-radiant Dicke ladder. This is shown in Fig.~\ref{fig:supp3} where we plot the total subradiant populations $P_\mr{sub}^{\mr{ss},\mu}(t_\mr{m},t_{\mr{m}}+\tau)$ due to two measurements resolved into the different Dicke ladder states with total angular momentum $J$. These promising results will be explored further as part of future work.

\subsection{Details regarding the numerical calculation of $P_\mr{sub}^{\mr{ss},\mu}$ and $t_\mr{sub}$}
In Figs.~2 and 3 of the main paper we have presented results of a numerical calculation for $P_\mr{sub}^{\mr{ss},\mu}$ and $t_\mr{sub}$ respectively. We present the details of these calculations briefly here. For the calculation of $P_\mr{sub}^{\mr{ss},\mu}$, the GKLS master equation (Eq.~(1) of the main paper) is evolved upto $t_\mr{m}$ within the bright Dicke subspace ($J=N/2$) to compute the state $\hrho(t_\mr{m})$. At time $t_\mr{m}$ the measurement probabilities and post-measurement states are calculated for the $\sigop_x$ or $\sigop_z$ measurement. As discussed above in Eqs.~\eqref{eq:postmeasz} and ~\eqref{eq:postmeasx} the resulting post-measurement states can be written within the $2N$ dimensional subspace of the measured emitter and the unmeasured emitters (in collective Dicke bright states with $\ket{J=(N-1)/2,M}_{N-1}$). We time evolve the post-measurement states for the two outcomes and calculate the large-time $t \gg t_\mr{m}$ (steady-state) population to be in the ground state $P_G^{\mr{ss},\mu} = 1-P_\mr{sub}^{\mr{ss},\mu}$ and the subradiant population. As evident from Fig.~2 of the main paper the numerical solution agrees perfectly with the analytical solution. The numerics to calculate the subradiant lifetime due to a single measurement at time $t_\mr{m}$ presented in Fig.~3 of the main paper follows a very similar structure. The only difference is that the time evolution upto $t_\mr{m}$, the post-measurement state, and the long-time evolution are all carried out in the full $2^N$ dimensional Hilbert space of $N$ TLEs.
\begin{figure}
\centering
\begin{overpic}[width = 0.95 \linewidth]{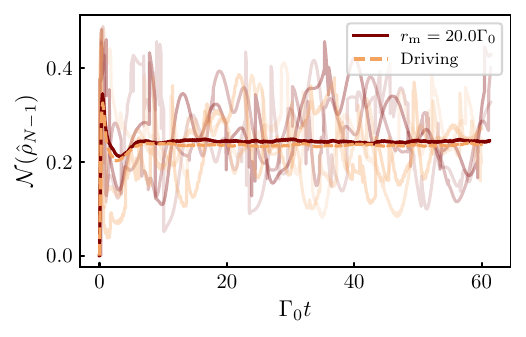}
\end{overpic}
\caption{Time evolution of the entanglement negativity of the unmeasured TLEs. The negativity $\mathcal{N}(\hrho_{N-1})$ is computed by averaging over 2000 MCWF trajectories (sample ones shown by translucent lines) for measurement on the $3^{\mr{rd}}$ TLE at the rate $r_{\mr{m}} = 20.0 \Gamma_0$ (maroon solid line) and for a Rabi drive with strength $\Omega_i =10 \Gamma_0$ (dashed orange line).}
\label{fig:supp8}
\end{figure}
\section{Derivation of the Zeno subspace master equation}
In this section, we consider the second protocol introduced in the main paper involving multiple $\sigop_x^i$ measurements of the $i^\mr{th}$ TLE of an array to prepare the unmeasured TLEs in a subradiant state and provide a detailed derivation describing the so called quantum Zeno limit where the number of measurements tends to $\infty$. The density matrix $\hrho(t)$ of the array of $N$ TLEs evolving under the Liouvillian $\mathcal{L}_N$ (see Eq.~(1) of the main paper) with $n_\mr{m}$ temporally equispaced non-selective $\sigop_x^i$ measurements (with the first measurement at $t_\mr{in} = 0$) can be written as
\begin{align}
    \hrho(t) = (e^{\mathcal{L}_N t/n_\mr{m}} \p_0)^{n_\mr{m}} \hrho(0) \label{eq:nmfinitemap},
\end{align}
where the quantum map $\mathcal{E}$ characterizing the nonselective measurement can be written as
\begin{align}
   \p_0[\bullet] = \sum_{k=\pm} \Pop_k \bullet \Pop_k, 
\end{align}
with the projection operators $\Pop_\pm$ for $\sigop^x_i$ taking the form
\begin{figure}
\centering
\begin{overpic}[width = 0.95 \linewidth]{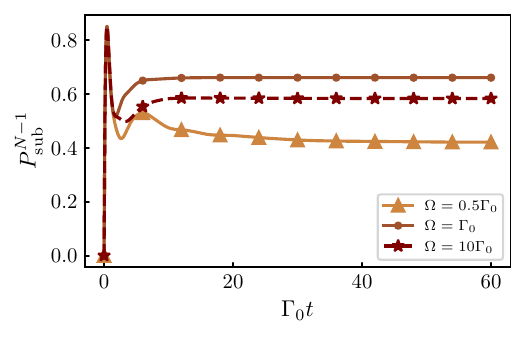}
\end{overpic}
\caption{Time evolution of the population of the subradiant Dicke state $P_\mr{sub}^{N-1}$ of $N-1$ undriven TLEs when the $i=3^\mr{d}$ emitter of an $N=7$ TLE array (spacing $d = 0.34\lambda_0$) interacting with a waveguide is driven with different values of Rabi frequency $\Omega_i$.}
\label{fig:supp9}
\end{figure}
\begin{align}
    \Pop_\pm = \ket{\pm_x}_i \bra{\pm_x} \otimes \bigotimes_{j \neq i}^N \mathbb{I}_j,\label{eq:Popplusminus}
\end{align}
where $\mathbb{I}_j$ denotes the identity operator for the $j^\mr{th}$ TLE. Using results from \cite{Burgarth2019generalized, Burgarth2020quantumzenodynamics}, the time-evolution in the quantum Zeno limit can be written as (noting $\p_0^2 = \p_0$)
\begin{align}
    \rhoop(t) = \lim_{n\rightarrow\infty} (\p_0 e^{\frac{t}{n}\mcl_N}\p_0)^n \rhoop(0) = e^{t\mcl_\mr{Z}}\p_0\rhoop(0),
    \label{eq:ZenoTimeEvolOp}
\end{align}
with the Zeno limit Liouvillian 
\begin{align}
\mcl_\mr{Z} = \sum_k \Pop_k \mcl_N \Pop_k.
\end{align}
Thus, as per Eq.~\eqref{eq:ZenoTimeEvolOp}, projection of the system dynamics to the Zeno subspace results in a modified GKLS master equation for the density matrix in the Zeno subspace $\p_0\hrho(t)$ of the form
\begin{align}
    \frac{d (\p_0\hrho(t))}{dt} = \mcl_\mr{Z} [\p_0\hrho(t)] \label{eq:GKSLZeno}.
\end{align}
Note that the above equation is a GKSL master equation crucially because the liouvillian operates on the projected density matrix $\p_0 \hrho(t)$ in the Zeno subspace \cite{Burgarth2019generalized, Burgarth2020quantumzenodynamics}. 

The calculation of $\mcl_\mr{Z}$ can be accomplished by dividing the terms of $\mcl_N$ into those involving the $i^{\mr{th}}$ emitter and those that do not \emph{i.e.}
\begin{align}
   & \mcl_\mr{Z}[\p_0\hrho] = \sum_{k=\pm} \Pop_k \left(-i\sum_{\mathclap{\substack{l\neq m \\ (l,m) \neq i}}}\Omega_{lm}\left[ \sigop_l^+\sigop_m^- , \p_0 \hrho \right]\right) \Pop_k \label{eq:zenoMEterm1}\\
&+ \sum_{k=\pm}\Pop_k \left(\sum_{(l,m) \neq i} \frac{\Gamma_{lm}}{2} \Diss{\sigop^-_l}{\sigop^+_m}[\p_0 \hrho] \right)\Pop_k \nonumber \\
&+  \sum_{k=\pm} \Pop_k \left(-i\sum_{l \neq i}\Omega_{li}\left[ \sigop_l^+\sigop_i^-+\sigop_i^+\sigop_l^- , \p_0 \hrho \right]\right) \Pop_k \label{eq:zenoMEterm2}\\
&+ \sum_{\mathclap{\substack{k=\pm\\ l}}}\Pop_k \left( \frac{\Gamma_{il}}{2} \left\{ \Diss{\sigop_i^-}{\sigop_l^+}[\p_0\rhoop]+ \Diss{\sigop_l^-}{\sigop_i^+}[\p_0\rhoop]\right\} \right) \Pop_k \nonumber
\end{align}
Since $\Pop_\pm$ commutes with all the operators in the first two lines above, we simply obtain the $\mathcal{L}_{N-1}$ for the $N-1$ unmeasured TLEs with the same form as introduced in Eq.(1) of the main paper. Performing the necessary simplifications in the third and fourth lines, we finally obtain
\begin{align}
    &\mcl_\mr{Z}[\p_0\hrho] = \mcl_{N-1}[\p_0 \hrho] -i \sum_{l \neq i} \frac{\Omega_{il}}{2}\left[\sigop^x_i \sigop^x_l,\p_0 \hrho  \right] \nonumber\\
   &+\frac{\Gamma_0}{2} \left( \mathcal{D}(\sigop^y_i)[\p_0 \hrho]  + \mathcal{D}(\sigop^z_i)[ \p_0\hrho]\right) \label{eq:LiouvZenoFull}\\
   & + \sum_{l \neq i} \frac{\Gamma_{il}}{4} \left ( \Diss{\sigop^x_i}{\sigop^+_l}[\p_0 \hrho] + \Diss{\sigop^x_i}{\sigop^-_l}[\p_0 \hrho] \right),
\nonumber
\end{align}
where for the sake of brevity, we have written $\mathcal{D}(\hat{A},\hat{B})$ as $\mathcal{D}(\hat{A})$ when $\hat{A} = \hat{B} = \hat{A}^\dagger$. From the above equation, we can next write down the dynamics of the unmeasured TLEs in the zeno subspace, \emph{i.e}., the time evolution of the reduced density matrix $\hchi = \mr{Tr}_i[\p_0\rhoop] = \hchi_+ + \hchi_-$, with $\hchi_\pm = \prescript{}{i}{\langle} \pm_x \vert \hrho \vert \pm_x \rangle_i$. Note that the evolution equation for $\hchi(t)$ is not in the GKLS form and is given by
\begin{align}
    \frac{d}{dt}\hchi &=  \frac{d}{dt}\hchi_+ +\frac{d}{dt}\hchi_- \label{eq:chievol},\\
    \frac{d\hchi_\pm}{dt} &= \mathcal{L}_{N-1} [\hchi_{\pm}] -\frac{\Gamma_0}{4} (\hchi_\pm - \hchi_\mp)  \label{eq:chievolmain}\\
&    -i \left[ \pm \left( \sum_{l \neq i} \frac{\Omega_{il}}{2} \sigop^x_l + \frac{\Gamma_{il}}{4} \sigop^y_l \right),\hchi_\pm \right ], \nonumber
\end{align}
which is precisely Eq.(3) of the main paper. 

We would like to note a couple of additional comments regarding the Zeno subspace effective master equation derived above. In the scenario of a permutation symmetric ensemble with $\Omega_{kl} = 0,\forall k,l$ and $\Gamma_{kl} = \Gamma_0 \forall k,l$, the effective equation for the $N-1$ unmeasured TLEs \eqref{eq:chievolmain} can again be written in terms of collective operators $\Sop^\mu_{N-1} = \sum_{l \neq i} \sigop^\mu_l, \mu=\{\pm,y \}$ and hence is permutation symmetric. Thus, we cannot get any occupation of the subradiant Dicke ladder in this scenario. Thus, for the second protocol to work, it is essential that we have permutation symmetry breaking. Secondly, let us consider the consequence of strongly measuring $\sigop^z_i$ instead of $\sigop^x_i$. The resulting evolution can be derived by replacing $\Pop_\pm$ in Eqs.~\eqref{eq:zenoMEterm1} and \eqref{eq:zenoMEterm2} by $\Pop_{e},\Pop_{g}$ with
\begin{align}
    \Pop_{e/g} = \ket{e/g}_i \bra{e/g} \otimes \prod_{j \neq i}^N \mathbb{I}_j\label{eq:Poppluseg}.
\end{align}
Due to the off-diagonal (in the $\ket{e}_i,\ket{g}_i$ basis) nature of the terms with $l \neq i$ in Eq.~\eqref{eq:zenoMEterm2} (both in the hamiltonian and dissipator), the dynamics of the $i^\mr{th}$ TLE and the unmeasured TLEs decouple. Thus, $\sigop^z_i$ measurement cannot drive the unmeasured TLEs to a steady Dicke subradiant states.

Apart from providing an intuitive picture of how strong measurement of a single TLE leads to subradiant steady state occupation for the unmeasured TLEs, the effective evolution in the zeno subspace Eq.~\eqref{eq:chievolmain} also provides a quick way to identify advantageous parameter regimes to obtain the highest subradiant steady state population. In Figs.~\ref{fig:supp4} and \ref{fig:supp5} we have calculated the subradiant population as a function of emitter spacing and the location $i$ of the measured TLE for $N=7$ and $N=5$ respectively. In Fig.~4 in the main paper, we have precisely chosen the maximum $d \sim 0.34 \lambda_0$ that leads to maximum subradiant population for the $N-1$ unmeasured TLEs.

\section*{Multiple Measurements - Additional Results on Excitation number, Entanglement Negativity}
We now present additional details supporting our central results for the second protocol presented in the main paper. A clear signature of steady-state subradiance is the presence of excitations in the ensemble. As we show in Fig.~\ref{fig:supp6}, for the scenario of $N=7$ emitters in a waveguide with the $i=3$ emitter is repeatedly measured, for large enough measurement rate $r_\mr{m}$, we see that the unmeasured TLEs clearly host a finite number of excitations confirming population of subradiant states in the long-time limit.

In addition to observables such as the population of subradiant Dicke states that are linear functions of the density matrix of $N-1$ emitters, we have also explored the behavior of purity and entanglement negativity and presented the behavior of the former in the main paper. For the sake of completeness and clarity, we would like clearly describe the procedure used in these calculations.
For both purity and entanglement negativity, we have computed the MCWF dynamics of $N$ TLEs interrupted by projective measurements of $\sigop^x$ of one of the TLEs at a constant rate $r_\mr{m}$. For each such trajectory, denoted by $\gamma$, we compute the reduced density matrix of the unmeasured TLES $\hrho_{N-1}^\gamma$ and calculate the corresponding purity
\begin{figure}
 \centering
 \begin{overpic}[width =\linewidth]{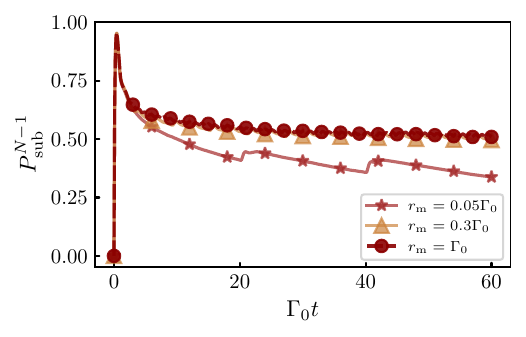}
 \end{overpic}
\caption{Subradiant population of the unmeasured emitters of an array of $N=10$ TLEs placed with separation $d =0.1\lambda_0$ inside a waveguide. The $\sigop_x^i$ of the emitter at location $i=1$ is measured repeatedly at the rate $r_\mr{m}$ starting from $t_{\mr{in}} = 0.25\Gamma_0^{-1}$.The population is computed by averaging over 5000 MCWF trajectories.}
\label{fig:supp10}
\end{figure}
\begin{align}
 \mathcal{P}^\gamma = \mathrm{Tr} ([\hrho_{N-1}^\gamma]^2) \label{eq:trajpurity},   
\end{align}
and the entanglement negativity
\begin{align}
    \mathcal{N}_\gamma = \frac{\left\|(\hrho_{N-1}^\gamma)^{\mr{T}_1} \right\|_1 -1}{2}\label{eq:trajnegativity},
\end{align}
where $\left\|(\hrho_{N-1}^\gamma)^{\mr{T}_1} \right\|_1$ denotes the trace norm of the partial transpose $(\hrho_{N-1}^\gamma)^{\mr{T}_1}$ with respect to half of the system of $N-1$ TLEs. Following this we average the two observables over many trajectories to determine the average purity $\mathcal{P} = \langle \mathcal{P}^\gamma \rangle_\gamma = \trajavg{\mathrm{Tr}(\hrho_{N-1}^2)}$ and entanglement negativity $\mathcal{N} = \langle \mathcal{N}^\gamma \rangle_\gamma$. Thus, this procedure which is also commonly used in the study of measurement induced phase transitions \cite{EhudAltman}, averages over the stochastic dynamics arising from both the quantum jumps in the MCWF trajectory and the local projective measurements. In Fig.~\ref{fig:supp7} (a,b) and Fig.~\ref{fig:supp8} we display the results for the average purity and negativity respectively and also show some a few representative trajectories in the background. These non-linear observables provide further evidence for the preparation of non-trivial steady-state with entanglement and significant purity. Moreover, as described in the main text, these observables provide a way to compare our protocol to the related strategy of strong local driving of one of the TLEs. In this case, while strong driving also produces a finite steady state probability to be in subradiant Dicke states (comparable to the strong measurement scenario) as shown in Figs.~\ref{fig:supp9} and ~\ref{fig:supp6}, the average purity reached in the Zeno limit with measurements is higher as shown in Fig.~\ref{fig:supp7} (same as Fig.~5 in the main paper). We also note that though the negativity is larger in the measurement based preparation as evident from Fig.~\ref{fig:supp8}, it does not show as much of an advantage as in the case of the average purity.

Finally, as we noted in the main paper, our procedure for the second protocol with repeated measurements on one emitter is to first optimize and find the configuration, in terms of array spacing $d$ and the location of the measured TLE $i$, that maximizes $P_\mr{sub}^{N-1}$ using the Zeno limit master equation. Given that the numerical calculation is performed in the full Hilbert space of $N$ TLEs, this procedure limits the number of TLEs to $N=7$ in the main paper. Nonetheless, as we show in Fig.~\ref{fig:supp10} with an unoptimized configuration with $N=10$ emitters, our results remain valid for larger values of $N$.


\bibliography{references.bib}